\documentclass[a4paper,  preprint, notitlepage, showkeys]{revtex4}

\usepackage{amsmath, amssymb, amsthm, mathtext}
\usepackage{graphicx}
\usepackage[caption=false]{subfig}

\newcommand{\tpr}{\boldsymbol{\otimes}}   
\newcommand{\dpr}{\boldsymbol{\cdot}}     

\newcommand{\vc}[1]{\boldsymbol{#1}}        
\newcommand{\mc}[1]{{\mathcal{#1}}}

\newcommand{\vcu}{{\vc{u}}}      
\newcommand{\vcx}{{\vc{x}}}      

\DeclareMathOperator{\codim}{codim}

\newcommand{\dudx}[2]{\frac{\partial{#1}}{\partial{#2}}}
\newcommand{\dudxf}[2]{\frac{d{#1}}{d{#2}}}
\newcommand{\ddudx}[2]{\cfrac{\partial^2{#1}}{\partial{#2}^2}}
\newcommand{\dhudhx}[2]{\frac{\partial_h{#1}}{\partial_h{#2}}}

\begin{document}
\title{
  On the diffuse interface models for high codimension dispersed inclusions
}
\author{ Elizaveta \surname{Zipunova} }
\email{e.zipunova@gmail.com}
\author{ Evgeny \surname{Savenkov} }
\email{e.savenkov@gmail.com}
\affiliation{Keldysh Institute of Applied Mathematics RAS, Moscow, Russia}
\date{\today}

\begin{abstract}
  Diffuse interface models are widely used to describe evolution of
  multi-phase systems of different nature. Dispersed
  ``inclusions'', described by the phase field distribution,
  are usually three dimensional objects. When describing elastic
  fracture evolution, elements of the dispersed phase are effectively
  2d objects. An example of the model which governs evolution of
  effectively 1d dispersed inclusions is phase field model for
  electric breakdown in solids. Phase field model is defined by
  appropriate free energy functional, which depends on phase field and
  its derivatives. In this work we show that codimension of the
  dispersed ``inclusion'' significantly restrict the functional
  dependency of system energy on the derivatives of the problem state
  variables. It is shown that free energy of any phase field model
  suitable to describe codimension~2 diffuse objects necessary depends on higher
  order derivatives of the phase field or need an additional
  smoothness of the solution~--— it should have first derivatives
  integrable with a power greater then two. To support theoretical
  discussion, some numerical experiments are presented.
\end{abstract}

\keywords{
  diffuse interface models; phase field; order parameter; electric
  breakdown}

\maketitle

\begin{acknowledgments}
  This work was supported by the Moscow Center for Fundamental and Applied Mathematics
  (agreement No.~075-15-2019-1623 with the Ministry of Education and Science of the Russian Federation).
\end{acknowledgments}

\clearpage

\section{Introduction}\label{sec:intro}
Phase-field (or order parameter) models constitute a powerfull and
theoretically sound framework for analysys of a broad class of
theoretical and applied problems in multi-phase
hydrodyanmics~\cite{lamorgese_2011,kim_2012,xu_2009,anderson_1997,santra_2020},
solid mechanics and fracture~\cite{ambati_2015}, material
science~\cite{provatas_2010}, solidification and phase
transitions~\cite{boettinger_2002, cartalade_2013, granasy_2014,
  gomez_2019},
crystal structures~\cite{emmerich_2012,asadi_2015,provatas_2007}
and many others topics.

The purpose if this class of models is, in general, a description of
the dynamics of certain ``inclusions''~--- elements of the dispersed phase immersed into homogeneous
medium. Such inclusions are usually elementary macroscopic
constituents (e.g., drops) of the dispersed phase of some multi-phase
system. Spatial distribution of the dispersed phase is described by
the so called phase field or order parameter field which is a smooth
function of time and spatial coordinate. Phase field function is
almost constant inside spatial domain occupied by the phase and
changes rapidly in the neighbourhood of the inter-phase
boundaries. For example, in the context of multi-phase hydrodynamics,
diffuse interface separates two immiscible fluids. Diffuse interface
is always of the finite width (hence the term) which is a model
parameter. Respectively, diffuse interface models has to provide
internal mechanisms which prevents exciding sharpening or spreading of
the diffuse interface during system evolution. This makes the diffuse
interface much more then just smoothed out contact discontinuity.

Comprehensive theoretical, applied and numerical analysis of the
diffuse interface models is a topic of huge amount of
literature. Either of these fields is a complex and complicated
subject~--- in particularly because of necessary non-linearity of such
models.

One of the central concepts of the diffuse interface models is the so
called ``sharp interface limit'' of the diffuse model. Roughly
speaking, the sharp interface counterpart of the diffuse interface
model is a model that is raised up when the diffuse interface width tends
to zero. For example, sharp interface limiting model for multi-phase
hydrodynamics is a free-boundary type model with sharp inter-phase
boundaries considered as two-dimensional surfaces complemented by the
respective interface conditions for conservative variables and their
fluxes. Transition form the diffuse interface model to the respective
sharp interface one usually is performed using asymptotic analysis
and/or~$\Gamma$-convergence
framework~\cite{caginalp_1991,elder_2001,abels_2014,abels_2018,rocca_2017,dellisola_2011,bourdin_2008}.

The connection between diffuse interface model and its sharp interface
counterpart is rather intimate. On the one side, for a number of diffuse
interface models the existence of it's physically meaningful sharp
counterpart gives theoretical and practical ground to the former
one. On the other side, thermodynamically consistent procedures to
derive constitutive relations for phase field models (using
Coleman-Noll procedure or Liu framework) can be applied 
only if provided with expression of the energy of the system defined
as a function of its primary termodynamical variables and their
gradients. This dependency can not be specified in more or less
precise way from the thermodynamical considerations~--- rather it is
the departure point of the consistent derivation~--- and has to be
postulated somehow else. In many cases sharp interface
limiting model (considered now as a primary one) is a source of
suitable thermodynamic relations, see, e.g.~\cite{lowengrub_2008}.

The topic of this paper is to consider such feature of the diffuse
interface models as spatial dimension (or codimension) of the
``inclusion'' which evolution is described by the phase field model.

For example, for diffuse interface models of multiphase hydrodynamics,
such ``inclusions'' are just droplets of the dispersed phase. Both
continuous (dispersion) phase and dispersive phase occupies effectively 3d spatial
domains. In this case the inclusion have dimension~3 and, hence,
codimension~0. The interphase boundary is effectively a 2d object with
codimension~1. For diffuse models of fractures the ``inclusion''
represents fracture mid-surface which is effectively of the
dimension~2 in 3d setting (and, respectively, of codimension~1).

Up to the authors knowledge, the only example of the phase-field model
which governs evolution of the codimension~2 (i.e., 1d
objects embedded in 3d medium) is presented in the
work~\cite{pitike_2014} which deals with phase field model for
electric breakdown of the non-conducting dielectric medium.

The goal of the present paper is to show that codimension of the
diffuse ``inclusion'' (or, more precisely, codimension of its sharp
interface counterpart) significantly restrict the functional
dependency of system's energy on the derivatives of the problem state
variables. In particular, keeping the form of this dependency as in the
lower codimension case, when considering models with higher
codimension ``inclusion'', can lead to the mathematically incorrect
problem setting.

Specifically, we show that a model suggested in~\cite{pitike_2014} can
not be used to describe electric breakdown channel as effectively
one-dimensional object. At the same time, description of the breakdown
channel as the three dimensional object doesn't need constructions
described in~\cite{pitike_2014}. Based on certain formal
considerations we propose corrected version of the phase field
electric breakdown model.

The aforementioned model~\cite{pitike_2014} is considered as a
specific example of the phase field model of ``codimension two''.
Nevertheless, we suppose that the presented considerations are of the general
interest and importance.

\section{Preliminary discussion}

The subject of this paper is the diffuse interface model for description
of electric breakdown process in solid dielectric medium suggested
in~\cite{pitike_2014,hong_2015} and further used in~\cite{cai_2017a,
  cai_2017b, cai_2019a, cai_2019b, cai_2019c}.  The
model~\cite{pitike_2014} is constructed as a formal generalization of
the well known phase field models for fracture evolution in elastic
medium. The derivation of the model is based on the formal analogy between
breakdown channel evolution and evolution of the fracture in an elastic
medium.

The motivation for such generalization is based on the following considerations.

In traditional theories of elastic fracture
mechanics~\cite{cherepanov_1967,anderson_1965}, the fracture
mid-surface is described as a sufficiently smooth two-dimensional
surface~$\Gamma$ embedded into the three dimensional space~--- i.e.,
as a geometrical object of codimension~$\codim\Gamma = 1$. Similarly,
electric breakdown channel~$\Lambda$ can be considered as a segment of
one-dimensional curve embedded into the three-dimensional space~---
being a geometrical object of codimension~$\codim\Lambda = 2$.

In both cases, the evolution of codimension~1 fracture mid-surface or
codimension~2 breakdown channel is governed by an appropriate internal
forces, of the elastic or quasi(electro)static nature, acting in the
medium. Besides this, for both fracture surface and breakdown channel,
it is possible to construct the so called~$J$-integral, which
describes energy release rate during fracture or channel evolution.

For fractures the concept of~$J$-integral is well known since
fundamental works of G.~Cherepanon~\cite{cherepanov_1967} and
J.~Rice~\cite{rice_1968}.  For breakdown channel evolution,
$J$-integral was derived in~\cite{suo_1993}. In this paper the
breakdown channel is described as one-dimensional conducting curve
which ``growth'' inside non-conducting medium. In both cases (for
elastic fractures and breakdown channel), $J$-integral can be used to
define fracture or channel growth criterion.

This formal correspondence between the processes (qualitatively, the
only formal difference between them is their \emph{codimension}) motivates authors
of~\cite{pitike_2014} to extend formally phase field fracture models
to phase field models for breakdown channel evolution.

To proceed, let us briefly discuss two main approaches for
construction of phase field models for fractures, see, e.g.~\cite{ambati_2015}
for comprehensive review.

The first approach, which is referred to as~``mechanistic''
hereinafter, is based on the framework of the variational fracture
theory, see~\cite{bourdin_2008},\cite[Chapter~3]{dellisola_2011}
and references there. It is based on the following constructions.

Consider spatial domain~$\Omega\subset\mathbb{R}^n$
($n = 1,2,3$), occupied by physically and geometrically linear
homogeneous elastic medium, to to which an external surface and
volumetric forces are applied.  Let~$\Gamma\subset\Omega$ be the
fracture mid-surface. In the two-dimensional setting~$\Gamma$ is
considered as segment of a smooth curve, in three-dimensional one~ ---
as a smooth surface with boundary. Within the variational fracture
theory (hereinafter we follow~\cite{bourdin_2008, dellisola_2011,
  borden_2014}) the energy of the system has the form of
\begin{equation}\label{eq:en:fr}
  \mathcal{J}
  =\mathcal{J}(\vcu,\Gamma)
  = \int\limits_{\Omega\setminus\Gamma} W(\nabla\tpr\vcu)\; d\Omega
  + \kappa\mc{H}^{n-1}(\Gamma),
\end{equation}
with~$\vcu$ being the displacement field, $\mc{H}^{n-1}(\Gamma)$~---
$(n-1)$-dimensional Hausdorff measure, $\kappa$ being the specific
energy attributed to the unit surface element of the fracture
mid-surface. The dimension of the Hausdorff measure is essential here:
fixing it to be equal to~$n-1$, one explicitly states that fracture is
the geometrical object of codimension~1.

The functional~\eqref{eq:en:fr} is the departure point for development
of the variational fracture theory. It can be shown that, under
certain assumptions which are of no importance now, the
displacements~$\vcu$ of the medium and the trajectory~$\Gamma(t)$ of
the fracture are minimizers of~\eqref{eq:en:fr} at each moment of
time. The details of the variational fracture theory are not
presented further, they are widely described elsewhere.

In the specified setting the analysis of the problem is complex~---
both from the theoretical and numerical points of view: to solve the
problem one need to be able to compute variations of~\eqref{eq:en:fr}
in respect to the fracture mid-surface configuration, that is, in respect to~$\Gamma$.

For this reason, it is more convenient to approximate the Hausdorff
measure in~\eqref{eq:en:fr} by the volume integral as (see~\cite{bourdin_2008, borden_2014})
\[
  \mc{H}^{n-1}(\Gamma)
  \approx \int\limits_\Omega \gamma_l(\phi,\nabla\phi)\;d\Omega,
\]
where volumetric approximation of the surface energy density reads
\begin{equation} \label{eq:en:fr:gamma}
  \gamma_l(\phi,\nabla\phi) = 
    \frac{1}{2l}\phi^2 + \frac{l}{2} \|\nabla\phi\|^2,
\end{equation}
with~$0 < l \ll 1$ being a small real valued parameter.

In this case the energy functional~\eqref{eq:en:fr} is \emph{approximated} by 
\begin{equation}\label{eq:en:fr:phi}
  \mc{J} \approx \mathcal{J}_l
  = \mathcal{J}_l(\vcu,\phi,\nabla\phi)
  = \int\limits_{\Omega} ((1-\phi)^2 + \epsilon ) W(\nabla\tpr\vcu)\; d\Omega
  +\int\limits_\Omega \kappa \gamma_l(\phi,\nabla\phi) \; d\Omega,
\end{equation}
where the order parameter (phase field) $\phi$ takes the
value~$\phi = 1$ on the fracture mid-surface~$\Gamma$ and~$\phi = 0$
in undamaged material.

In~\eqref{eq:en:fr:phi},~$0<\epsilon \ll 1$ is a small real valued
parameter which prevents degeneracy of the functional
when~$\phi=1$. Usually it is chosen to be strictly positive~---
although it is known, that even with vanishing~$\epsilon$ , the
functional~\eqref{eq:en:fr:phi} is not degenerate,
see~\cite{braides_1998}. It can be shown that~\eqref{eq:en:fr:phi}
$\Gamma$-converges to the functional~\eqref{eq:en:fr}
when~$l\to 0$.

If the fracture mid-surface~$\Gamma$ is prescribed and displacement field is vanishing (i.e., the median is not loaded), then it is easy to
show that the distribution of~$\phi$ in space is
defined as the minimizer of the functional
\begin{equation}\label{eq:en:fr:phi:2}
  \mathcal{\Tilde{J}}_l
  =\mathcal{\Tilde{J}}_l(\phi,\nabla\phi)
  = \int\limits_\Omega \kappa \gamma_l(\phi,\nabla\phi) \; d\Omega,
\end{equation}
subjected to the boundary conditions
\begin{equation}\label{eq:fr:phi:bc}
  \phi|_\Gamma = 1;\quad \phi\to 0,\; \vcx\to\infty.
\end{equation}

The corresponding Euler-Lagrange equations read:
\begin{equation}\label{eq:fr:phi}
  -\Delta\phi + \frac{1}{l^2}\phi = 0, \quad \vcx\in \Omega.
\end{equation}
It can be shown that in 1d case($n=1$, $\Gamma = \{x_0 = 0\}$), the
solution of the latter equation is
\[
  \phi(x) = \exp(-|x|/l)
\]
and decreases exponentially as the distance from a point in space to
the fracture mid-surface increases. In the multidimensional case, this
property also holds. Now it can be seen that parameter~$l$ in the
expression for~$\gamma_l$ defines the width of the diffuse fracture.

The equation~\eqref{eq:fr:phi} above is the model one.  In the
complete formulation of the problem, when the path~$\Gamma(t)$ of the
fracture evolution is not known a priori, this equation is solved in
the whole domain~$\Omega$ with a source term which depends on the
local elastic energy of the medium.  The points in space at which the
value~$\phi = 1$ are assumed to be the points of the fracture. As a
result, a new fracture surface (that is, a set of points
where~$\phi$ takes the value~1) is formed where, for example,
sufficiently large tensile elastic stresses appears.

The second approach, which will be further called ``thermodynamic'',
is based on a-priory definition of the form of energy
functional. With this definition, one postulates or derives, within the rational
thermomechanics framework, the equations defining the fracture
evolution and the state of the surrounding medium. These equations
have a standard form, typical for a phase-filed models~--- in
particular, the evolution of the order parameter~$\phi $ is usually
described by a classical equation of the Allen-Kahn type.

Note that both approaches are closely related in the sense that using
a mechanistic approach, thermodynamically consistent models can be
derived. However, their correctness must be proved a-posteriory~--- in
contrast to the thermodynamic models, which are correct by
construction. Neither of two methods is ``more correct'' or
``less correct''. Indeed, in thermodynamic models it is necessary to
define, from the very beginning, the form of energy of a medium with a
fracture~--- which cannot be predicted by purely phenomenological
considerations. An understanding of how this energy can be defined
comes from considering ``mechanistic'' models.

An issue which is not explicitly discussed in the literature~--- and which is
the topic to which the present work is devoted,~--- is the
following. The choice of an expression for the energy
density~$\gamma_{l} $ and the corresponding
functional~\eqref{eq:en:fr:gamma} is not unique and has a certain
arbitrariness. Nevertheless, definitely, the energy has to be chosen in such a way,
that a minimization problem statement for the functional
\begin{equation}\label{eq:cap:1}
  \Tilde{\mathcal{J}}_l = \int\limits_\Omega\gamma_l\; d\Omega \to \min.
\end{equation}
subjected to the boundary conditions~\eqref{eq:fr:phi:bc}
is mathematically correct.

In the case when the ``diffuse'' object is a part of the surface in
three-dimensional space,~--- that is, its sharp counterpart has
codimension~1~--- such questions do not arise since the
boundary value problem is usually posed in the domain which boundary
has its natural codimension~1 (i.e., it is a two-dimensional surface in
three-dimensional case or one-dimensional in two-dimensional case)~---
and in this sense it does not differ from the
boundary~$\partial\Omega$ of the domain~$\Omega $,
$\dim \Gamma = \dim \partial \Omega = 2 $. Obviously, if a diffuse
object has larger codimension, the issue described above should be
taken into account.

Consider now the diffuse interface model of the breakdown channel,
presented in~\cite{pitike_2014}.  Essentially, it is constructed as a
formal generalization of the fracture diffuse model. In particular, it
is implicitly assumed that the behavior of the system is described by
the energy functional, which has the form (cf.~\eqref{eq:en:fr})
\begin{equation*}
  \mathcal{J}
  =\mathcal{J}(\Phi,\Gamma)
  = \int\limits_{\Omega\setminus\Lambda} W(\nabla\tpr\Phi)\; d\Omega
  + \kappa\mc{H}^{n-2}(\Lambda),
\end{equation*}
where~$\Phi$ is electric field potential, $\mc{H}^{n-2}(\Lambda)$
stands for $n-2$-dimensional Hasudorff measure, $\kappa$ is equal to
the specific energy assigned to the unit length of the breakdown
channel. The dimension of the Hausdorf measure is essential here: this
time it is equal to~2 (in 3d case), which explicitly says that breakdown channel is
a 1d object embedded in 3d space.

The key point now is the question of how should the corresponding part
of the energy (that is,~$\gamma_l$) of the system be specified in order to
approximate in the correct way the values of
\begin{equation}\label{eq:cap:2}
  \mc{H}^{n-2}(\Gamma)
  \approx \int\limits_\Omega \gamma_l(\phi,\nabla\phi)\;d\Omega,
\end{equation}
subjected to the boundary conditions~\eqref{eq:fr:phi:bc}.  The answer
to this question essentially depends on the codimension of~$\Lambda $.
As it will be shown below, the expression~\eqref{eq:en:fr:gamma}
cannot be used if~$\Lambda$ has codimension 2, i.e., if~$\Lambda$ is a
curve in 3d case or a point in planar case.

Note that the expressions~\eqref{eq:cap:1} and~\eqref{eq:cap:2} with
the boundary conditions~\eqref{eq:fr:phi:bc} essentially defines the
capacity of the set~$\Lambda$ relative to~$\gamma_l $. Therefore,
the question of the correctness of the considered minimization
problems is closely related to the theory of the capacity of sets~---
in particular, with the question of whether the capacity of a set of a
given codimension is positive~--- informally, that is the criteria to
check if the set~$\Lambda$ supports definition of the boundary
condition of the given type.

\section{Formal description of the model from~\cite{pitike_2014} }

Let us briefly outline the phase field model proposed in~\cite{pitike_2014}
to describe electric breakdown channel propagation in a solid dielectric.

Consider a bounded domain~$\Omega$ occupied at the initial
time~$t = 0 $ with a solid dielectric with electric
permeability~$\epsilon = \epsilon_0 (\vcx, t)$.  During the
electrical breakdown, the formation of a breakdown channel occurs. The physical breakdown channel can be described as a cylindrical domain of small radius.  By
analogy with fractures in an elastic medium, this domain is considered
as a damaged domain with alternative properties.  According to the
diffuse interface method, it is assumed that the spatial distribution of
the damaged material is described by at least continuous scalar
function (phase field or order parameter)~$\phi = \phi(\vcx, t)$,
defined in~$\Omega $. The range of this function is a
segment~$[0,1]$, its value~$\phi = 0 $ corresponds to the medium in
the breakdown channel, the value~$\phi = 1 $~--- to the undamaged
medium.  The values~$\phi \in (\epsilon, 1- \epsilon)$,
$\epsilon \ll 1 $ correspond to the diffuse boundary separating the
breakdown channel ($ \phi = 0 $) from the intact medium
($ \phi = 1 $). Equations defining the evolution of the order
parameter~$ \phi$ are chosen so that the channel domain is a small
tubular neighborhood of a curve in space that corresponds to the axis
(middle curve) of the breakdown channel.  The effective diameter of
this tubular neighborhood in equilibrium state is defined by a model
parameter.

Assuming the breakdown channel be an ideal conductor, its electric
permittivity is infinitely large. In the model, in accordance with the
ideas of the diffuse interface approach, the permittivity is assumed
to have finite, but very large values.  Specifically, it is defined as
\begin{equation}
  \label{eq:eps}
  \epsilon = \epsilon[\phi](\vcx,t)
  = \frac{\epsilon_0(\vcx)}{f(\phi(\vcx,t)) + \delta}.
\end{equation}
Here~$\epsilon_0=\epsilon_0(\vcx)$ is electric permittivity of the
undamaged medium; $f=f(\phi)$ is the so called interpolation (or
degradation) function; $0<\delta\ll 1$ is a small positive real valued
regularizing parameter, which prevents degeneracy
of~\eqref{eq:eps} at~$\phi=0$.

The main role of the function~$ f $ is the interpolation of the medium
properties and parameters of the model between the damaged and
undamaged values, which corresponds to the ``pure'' phases identified
by $\phi = 0$ and $\phi = 1$, see~\cite{sargado_2018}. In~\cite{pitike_2014},
it is chosen as~$ f(\phi) = 4 \phi^3 - 3 \phi^4$.

Since the development of the breakdown channel is essentially slower
than the speed of electromagnetic waves propagation in the medium, it
is assumed that the energy of the magnetic field can be neglected. As
a result, the problem is considered in quasi (electro) static setting
and distribution of the electric field is potential.

As a result, within the formal analogy with phase field fracture models, the
following dependency of free energy on state variables and their
derivatives is postulated in~\cite{pitike_2014}:
\begin{equation}
  \label{eq:en_vol}
  \Pi = \int\limits_\Omega \pi\, d\Omega,\quad  \pi = \pi (\Phi,\phi,\nabla\phi),
\end{equation}
\begin{equation}
  \label{eq:en}
  \begin{aligned}
    \pi &= -\frac{1}{2}\vc{E}\dpr\vc{D} + \Gamma \frac{1-f(\phi)}{l^2} + \frac{\Gamma}{4}\nabla\phi\dpr\nabla\phi = \\
        &= -\frac{1}{2}\epsilon[\phi]\nabla\Phi\dpr\nabla\Phi + \Gamma \frac{1-f(\phi)}{l^2} + \frac{\Gamma}{4}\nabla\phi\dpr\nabla\phi.
    \end{aligned}
\end{equation}
with~$\vc{E} = -\nabla\Phi$ being the electric field, $\Phi$ being its
potential, $\vc{D} = \epsilon \vc{E}$ being the electric induction;
$\Gamma$ is specific energy per unit length of the channel and $l$
defines its effective radius.

The first term in~\eqref{eq:en} is the electrostatic energy of the
medium.  Other terms are ``phase field part'' of the energy and are
chosen by the authors of~\cite{pitike_2014} formally~--- they coincide
with that for the energy used in the diffuse fracture models. The primary
state parameters of the model are~$\Phi$, $\phi$ and~$\nabla \phi$.

The system of equations that describes evolution of the system in the
non-stationary case is postulated in the form
\begin{subequations}
  \label{eq:el}
  \begin{gather}
    \label{eq:el:1}
  \frac{\delta \pi}{\delta \Phi} = 0,\\
  \label{eq:el:2}
  \frac{1}{m}\dudx{\phi}{t} = - \frac{\delta \pi}{\delta \phi}.
\end{gather}
\end{subequations}
Above, the first equation~\eqref{eq:el:1} describes the distribution
of electric potential~$\Phi $. The second one is the simplest equation
describing the kinetics of the order parameter. The parameter~$m>0$ is
phenomenological parameter called mobility with the meaning of the rate of
change of a given quantity under the action of a applied unit force.
The equation~\eqref{eq:el:2} formalizes the empirical assumption that
the deviation of the spatial distribution
of the order parameter from the equilibrium state evolves so that to compensate the deviation.

In the expanded form equations~\eqref{eq:el} read:
\begin{gather}
  \nabla\dpr \left( \epsilon[\phi]\nabla\Phi \right) = 0, \\
  \label{eq:AllenCahn}
  \frac{1}{m} \dudx{\phi}{t} = \frac{1}{2}\epsilon'(\phi) \nabla\Phi\dpr\nabla\Phi +
  \frac{\Gamma}{l} f'(\phi) + \frac{1}{2}\Gamma \Delta\phi,
\end{gather}
where~$(\cdot)' \equiv (\cdot)'_\phi$
and dependency $\epsilon = \epsilon(\phi)$ is defined by~\eqref{eq:eps}.

The first equation above is the equation for the electric potential
with a dielectric permittivity depending on the distribution of the
order parameter. The second equation has the form of an Allen-Cahn type
equation which describes the spatial and temporal evolution of the
order parameter.

The equilibrium state of the medium with free energy defined
by~\eqref{eq:en} is defined by the conditions of vanishing variations
of~\eqref{eq:en} in~$\Phi $ and~$\phi$. The corresponding
Euler-Lagrange equations are formally correspond to the
equations~\eqref{eq:el} with~$1/m \to 0 $ and have the form of:
\begin{equation}\label{eq:el:eq} 
  \nabla\dpr \left( \epsilon[\phi]\nabla\Phi \right) = 0,\quad
  \frac{1}{2} \epsilon'[\phi] \nabla\Phi\dpr\nabla\Phi + \frac{\Gamma}{l^2}f'(\phi) +
  \frac{\Gamma}{2}\Delta \phi = 0.
\end{equation}

To determine the structure of the diffuse interface described by the
model~\eqref{eq:el:eq}, consider an unbounded domain in which the axis
of the breakdown channel is a smooth curve~$\Lambda$. Assuming the
vanishing electric field in the medium, the order parameter~$\phi$
satisfies the following boundary conditions:
\begin{alignat*}{2}
  &\text{for } \vcx\in\Lambda:& \quad & \phi(\vcx)  = 0, \\
  &\text{for } \vcx\to\infty:& \quad & \phi(\vcx)  \to 1,
\end{alignat*}
and the governing equation: 
\begin{equation}\label{eq:sing:1}
  \frac{\Gamma}{l^2}f'(\phi) +  \frac{\Gamma}{2}\Delta \phi = 0
\end{equation}
defined in~$\mathbb{R}^3\setminus\Lambda$.

The solution to this equation describes the distribution of the order
parameter, which corresponds to the straight conductor of infinitely
small diameter.  Here, a conductor is understood as a set of points in
space with~$ \epsilon = +\infty $, which corresponds to the
value~$\phi = 0 $.

Define cylindrical coordinate system~$\mc{O}r\theta{z}$, with~$ r $
being the distance from a point in space to the axis~$\mc{O} z $ and
$\theta $ being the polar angle.  In what follows, we will assume that
the breakdown channel coincides with the~$\mc{O} z $ axis, that
is,~$\Lambda = \mc{O} z $ and the order parameter
distribution~$ \phi $ depends only on the radius, i.e.,
$ \partial \phi / \partial \theta = \partial \phi / \partial z = 0 $
and~$ \phi = \phi (r) $.

In~\cite{pitike_2014} it is stated that axisymmetric distribution of
the order parameter satisfy the equation
\begin{equation}
  \label{eq:wrong:1}
  f'(\phi) + \frac{l^2}{2}\frac{d^2 \phi}{d r^2} = 0
\end{equation}
with boundary conditions
\begin{equation}\label{eq:bc}
  \phi|_{  r = 0} = 0; \quad \phi|_{r\to+\infty}\to 1.
\end{equation}
Integration of~\eqref{eq:wrong:1} leads to the equation
\begin{equation}
  \label{eq:wrong:2}
  \dudxf{\phi}{r} = \frac{2}{l}\sqrt{1 - f(\phi)}.
\end{equation}
In~\cite{pitike_2014}, 
equations~\eqref{eq:wrong:1} and~\eqref{eq:wrong:2} are equations~(13) and~(14).

\section{Analysis of the model~\cite{pitike_2014}}\label{sec:analysis}

The model described in the previous section is the only model known to
the authors in which the diffuse interface model is used to describe
an object with codimension~2. Although the method of its construction
is the ``mechanistic'' one, it is relatively common and widely used.

Nevertheless, its particular implementation in~\cite{pitike_2014} is
not completely correct. More precisely, the form of free energy used
to describe a (codimension~1) ``diffuse'' fracture cannot be used to
describe a (codimension~2) ``diffuse'' breakdown channel. As will be
shown below, this is due to the fundamental mathematical properties of
the corresponding expression for the free energy.

First, the equation~\eqref{eq:wrong:1}, which is
positioned~\cite{pitike_2014} as an equation describing an
axisymmetric distribution of the order parameter, is incorrect.

Indeed, consider the primary equation~\eqref{eq:sing:1} for the phase
field distribution in the three-dimensional domain, which contains an
infinitely long breakdown channel.

Let us remind definition of the Laplace operator in the
cylindrical coordinates~$\mc{O}r\theta{z}$,
\begin{equation}\label{eq:lap_cyl}
  \Delta \phi = \frac{1}{r}\dudx{}{r}\left(r\dudx{\phi}{r}\right) +
  \frac{1}{r^2} \ddudx{\phi}{\theta}  + \ddudx{\phi}{z}.
\end{equation}
For one dimensional axially symmetric case
($\partial \phi/\partial z = \partial \phi/\partial \theta = 0 $,)
considered in~\cite{pitike_2014}, it followos from~\eqref{eq:sing:1}
and~\eqref{eq:lap_cyl} that
\begin{equation}\label{eq:sing:1:a}
  \frac{1}{l^2}f'(\phi) +
  \frac{1}{2} \frac{1}{r}\dudx{}{r}\left(r\dudx{\phi}{r}\right) = 0,
\end{equation}
or, expanding derivatives, 
\begin{equation}\label{eq:sing:2}
  \frac{1}{l^2}f'(\phi) +
  \frac{1}{2} \left( \frac{1}{r} \dudx{\phi}{r} + \ddudx{\phi}{r} \right) = 0.
\end{equation}

First of all let us note that equations~\eqref{eq:sing:1:a}
and~\eqref{eq:sing:2} differs form~\eqref{eq:wrong:1}
and~\eqref{eq:wrong:2}, given in~\cite{pitike_2014} as equations~(13)
and~(14).  It is easy to notice that~\eqref{eq:wrong:1}
and~\eqref{eq:wrong:2} are correct, but only for one dimensional
\emph{planar} case and not for the axysimmetric one as it is stated
in~\cite{pitike_2014}.

This error would be rather technical one if not for the following observation.

To construct a solution to the one-dimensional
equation~\eqref{eq:sing:2}, the boundary conditions~\eqref{eq:bc} with
respect to solution \emph{values} has to be accounted at point~$r = 0$
and~$r \to + \infty$.

It is well known that setting the boundary condition at~$r = 0 $
(which is a \emph{point set}) on the value of the solution to
determine axisymmetric solutions of the \emph{second order} equation
of the type~\eqref{eq:sing:1} is incorrect. The rationale for this
statement is the presence of the so-called theorems on ``removable
isolated singularities'' in the theory of PDEs.  In particular, such a
result is well known in the theory of harmonic functions. Simply
speaking it states that if a function is bounded and harmonic outside
any arbitrarily small neighborhood of the given point, then it can be
extended to this point as harmonic function. This means that the value
of such function can not be defined arbitrarily at a single point.

The equation~\eqref{eq:sing:1} considered here is not the Laplace
equation, but is the second order semi-linear elliptic equation with
Laplacian as a leading term.  Consider it in the two-dimensional case
(that is,~$\partial \phi / \partial z = 0 $).  The requirement for
the axisymmetry of the solution is no longer will be needed.
For~\eqref{eq:sing:1} the required result is given
in~\cite{cirstea_2007,veron_1986} and can formulated as follows (see
also~\cite{hirata_2011, hirata_2014}):

  Consider equation~\eqref{eq:sing:1} defined in the
  domain~$\Omega\subset\mathbb{R}^2$. Let~$\phi = \phi(x)$ be its
  solution in~$\Omega\setminus\omega(\vcx_0)$ with~$\omega(\vcx_0)$
  being an arbitrarily small neigborghood of the
  point~$\vcx_0\in\Omega$. Then $\vcx_0$ is removable singularity~---
  that is,~$\phi = \phi(\vcx)$ can be extended to~$\Omega$ as the
  solution of~\eqref{eq:sing:1} \emph{iff}~$\phi$ growth not faster
  than~$\mu(\vcx) = \ln(1/\|\vcx-\vcx_0\|)$, i.e.,
  $\lim_{\vcx\to \vcx_0} {\phi(\vcx)}/{\mu(\vcx)} = 0$

The boundary condition~$\phi(\vcx_0) = \phi_0$ with
finite~$\phi_0$ at the point~$\vcx_0 = 0$ ensures fulfillment of these
conditions. Hence,~$\phi(\vcx)$ satisfies equation~\eqref{eq:sing:1}
in whole domain~$\Omega$ with boundary conditions defined at its
outer boundary.  For axisymmetric problem with~$\Omega$ being a disk
of the given radius centered at~$\vcx_0 = 0$ and boundary
conditions~\eqref{eq:bc} at~$r = R$ or $r \to+\infty$, such solution
is the trivial one, i.e., $\phi(\vcx) = 1$.

From here it follows that overall setting considered
in~\cite{pitike_2014} is not mathematically correct~--- since point
boundary conditions can not be set for semilinear 2-nd order PDEs of
the considered form.

Thus, the results presented in~\cite{pitike_2014}, in particular,
equation~(13), (14) and the solution in figure~2 (here the references are given
according to~\cite{pitike_2014}) are not correct~--- they describe the
\emph{planar} case, which corresponds to the diffuse interface models
for fractures and is mathematically sound.

In this case, actually covered in~\cite{pitike_2014}, posing boundary
conditions for the values of solution at~$r = 0$ is possible~ --- in
the planar case the set of points $r = 0$ is a line on a plane (in a
two-dimensional case) or a plane in space (in~3d setting).  In other words,
the equation~\eqref{eq:sing:1} in this case is considered in the
half-space~$0 < x_1 < +\infty $, where~$ x_1 \equiv r $.

\begin{figure}[t!]
  \centering
  \includegraphics[width=0.75\textwidth]{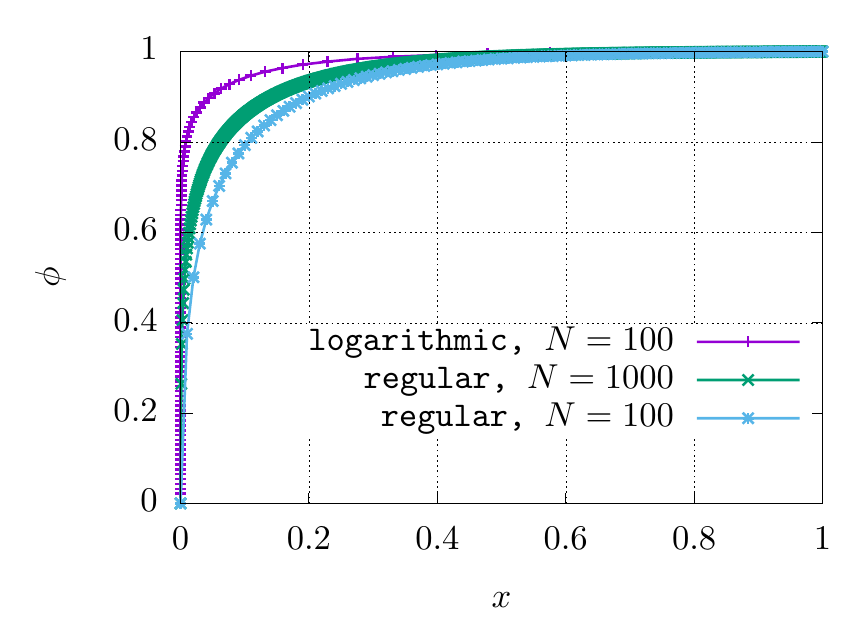}
  \caption{Lack of numerical convergence of the numerical solution of eq.~\eqref{eq:sing:1:a}.
  }\label{fig:conv}
\end{figure}

Note that the equation~\eqref{eq:sing:1:a} can be formally approximated
by a suitable difference scheme in the domain~$r \in [0, R] $ with a
given Dirichlet boundary conditions at~ $r = 0, R$.  In this case, the
numerical solution will have a qualitative form, presented in figure~2
in~\cite{pitike_2014}. However, when refining the mesh step
size, there will be observed complete absence of grid convergence: with
mesh step size going to zero, the solution will asymptotically tends
to 1 at each point except for~$ r = 0 $. In
other words, the derivative of the numerical solution at the point
$ r = 0 $ will be tend to zero at the point $r=R$ and to infinity
at~$r=0$, see figure~\ref{fig:conv}.
However, as can be easily shown, this is not the case for the solution
of the equation~\eqref{eq:wrong:2}. Indeed, for~$ r \to 0 $ we
have~$\phi\to 0 $ and from the equation~\eqref{eq:wrong:2} it
follows that~$ d \phi / dr \to 1 $. Similarly, if the equation~\eqref{eq:wrong:2}
is solved in the region~$0<r<R < +\infty$
with the boundary condition~$\phi(r = R) = 1$, then for~$r\to R$
we have~$\phi \to 1 $ and, as a consequence, $d\phi/dr \to 0 $.

It can be assumed that the lack of numerical convergence was not
discovered by the authors of~\cite{pitike_2014} due to the fact that
authors solve numerically not the ``primary''
equation~~\eqref{eq:sing:1:a} or~\eqref{eq:sing:2}, but the
equation~\eqref{eq:wrong:2}, which corresponds to the planar, not
axisymmetric case.

\section{Corrected model}

Thus, the setting of the boundary conditions for equations~\eqref{eq:sing:1} is not mathematically possible on the codimension~2
set~$\Lambda$, which models the breakdown channel. To correct the
model one need to modify expression~\eqref{eq:en} for the energy of
the system in such a way, that for the corresponding Euler-Lagrange
equation~\eqref{eq:sing:1}, definition of the Dirichlet boundary
condition on~$\Lambda$ with $\codim \Lambda = 2 $ would be
possible.

The easiest way to see how this can be done is to consider the
equation~\eqref{eq:sing:1} in weak (variational) setting.  Consider
the equation~\eqref{eq:sing:1} in the two-dimensional
domain~$ \Omega $, which is assumed to be simply connected and without
``punctured'' points.  In this case, the weak statement of the
problem~\eqref{eq:sing:1} in the two-dimensional region~$\Omega\subset\mathbb{R}^2$ has the form: find
a function~$ u \in V_0 $ such that
\begin{equation}
  \label{eq:var:1}
  a(u,v) + (f'(u), v) = 0,\quad
  a(u,v) = \int\limits_\Omega\nabla u \dpr  \nabla v\, d\Omega,\;
  (f(u),v) = \int\limits_\Omega f(u)v\, d\Omega,
\end{equation}
for an arbitrary test function~$v\in V_0$. Here the functional
space $V_0\subset V$ consist of functions from~$V$ vanishing
on~$\partial\Omega$, $V$ is a space of sufficiently smooth functions
defined in~$\Omega$. Here ``sufficiently smooth''  means that function
from~$V$ have finite energy norm induced by the bilinear
form~$a(\cdot,\cdot)$.

For the second order elliptic problem~\eqref{eq:var:1} the natural
smoothness is~$ V = W_2 ^ 1 (\Omega) $, that is, the Sobolev space of
functions which are $L_2$-integrable and have~$L_2$-integrable
gradient. Accordingly, the space $ V_0 = W ^ 1_ {2,0} (\Omega) $.

Due to the well-known trace theorems (see~\cite{sobolev_1988,
  adams_2003}), for functions $ v \in H ^ 1 (\Omega) $ the trace
operator is defined only for the sufficiently smooth manifolds of
codimension~1, i.e., surfaces in 3d. For manifolds of
codimension~$ 2$~--- which is of the interest here~--- the trace
of functions $ v \in H ^ 1 (\Omega) $ cannot be defined. From here
it is clear that boundary conditions can not be defined for a function
from~$H^1(\Omega)$ except the boundary is a codimension~1 manifold.

Consider Sobolev space~$W^q_p(\Omega)$, $\Omega\subset\mathbb{R}^n$,
$n = 3$, which can be defined as
\[
  \|u\|_{W^q_p(\Omega)}^p
  = \sum\limits_{|\alpha|\leqslant q} \|D^\alpha u \|^p,\quad
  \|\cdot\| = \|\cdot\|_{L_p(\Omega)},
\]
where~$\alpha = (\alpha_1,\alpha_2,\ldots,\alpha_n)$ is multi-index
\[
  D^\alpha = \frac{\partial^{|\alpha|}}{\partial x_1^{\alpha_1}x_2^{\alpha_2}\ldots x_n^{\alpha_n}},\quad
  |\alpha| = \alpha_1 + \alpha_2 + \cdots+\alpha_n.
\]
According to the embedding theorems, one have
\[
  W^q_p(\Omega) \subset C^{r,\alpha}(\Omega), \quad |\alpha|\leqslant q,
\]
if
\[
  n < p q,\quad \frac{1}{p} - \frac{q}{n} = -\frac{r+\alpha}{n},
\]
where~$C^{r,\alpha}(\Omega)$ is the respective H\"{o}lder space.

This means that functions for which the value of~$pq$ is sufficiently large,
will be at least continuous. In the  considered case, one have~$n = 3$, so that
\[
  pq > 3,\quad  \frac{1}{p} - \frac{q}{3} = -\frac{r+\alpha}{n}.
\]

Restricting ourselves to the minimal natural values~$p$ and~$q$
satisfying the last inequality, we have~$pq = 4 $, whence:
\begin{align*}
  p = 2, \; q = 2: \quad r = 0,\alpha = 1/2, \\
  p = 4, \; q = 1: \quad r = 0,\alpha = 1/4.
\end{align*}

This means, that the traces of the order parameter~$ \phi $ will be
correctly defined on manifolds of codimension~2 if from the finiteness
of the values of the functional~$ \pi $ it will follow
that~$ \phi \in W_p ^ q (\Omega) $ with~$p$ and~$ q $ given above~---
that is for~$\phi\in W_2^q(\Omega)$ for~$q\geq 2$
and/or~$\phi\in W_p^1(\Omega)$ for~$p\geq 4$.

In the first case the energy has to include at least the term 
\[
  \pi_{2,2} = \int\limits_\Omega (\Delta \phi)^2  \; d\Omega.
\]
Hence, the corresponding Euler-Lagrange equation will have the
polyharmonic term of the form~$\Delta^2 \phi$.  In the second case,
the energy has to include at least the term
\[
  \pi_{1,4} = \int\limits_\Omega \frac{1}{p}\|\nabla \phi\|^p  \; d\Omega,\quad p\geqslant 4.
\]
Hence, the corresponding Euler-Lagrange equation will have the so
called~$p$-Laplacian term
~$\Delta_p\phi\equiv\nabla\dpr \left( \|\nabla\phi\|^{p-2}\nabla\phi\right)$.
In what further we consider only the case of~$p=4$.

Note that the boundary value problems for the polyharmonic
equation~$\Delta ^ k u = f $, $ k = 2,3, \ldots $, with boundary conditions
defined at
\[
  \Gamma = \Gamma_0 \cup \Gamma_1 \cup \Gamma_2 \cup \ldots,
\]
with~$\Gamma_k $ being a manifold of codimension~$k$,
$ \dim \Gamma_k = k $, $ \codim \Gamma_k = n-k $, $ n = \dim \Omega $,
$\Gamma_0\equiv\partial\Omega$, are studied in~\cite{sobolev_1988, oleinik_1996a, oleinik_1996b, sternin_1964}.

Correctness of the boundary value problems for the quasilinear elliptic equation with~$ p $-Laplacian is considered in~\cite{lewis_2018}.

Note that the main results of the theory of boundary value problems
with boundary conditions, given on manifolds of high codimension, was
basically obtained in the context of the theory of capacity of
sets. Namely, one can often assume that a manifold of high codimension
supports definition of the boundary conditions if the corresponding
manifold has positive capacity with respect to problem’s
operator. Currently, the theory of (degenerate) partial differential
equations in domains whose boundaries include sets of high
codimension, is a new and actively developed topic of the ODE theory ,
capacity theory and geometric measure theory, see, e.g.~\cite{david_2020} and references therein.

Thus, if the kinetic equation~\eqref{eq:el:2} describing the
dynamics of the order parameter will include a polyharmonic operator
and/or $p$-Laplacian term, then for its solution the setting of
boundary conditions on a manifold of codimension~1 will make
sense. This is the hint to correct the model from~\cite{pitike_2014}.


The formal generalization of the energy which ensures the correctness
of the diffuse interface model of the breakdown channel has the form:
\begin{equation*}
  \Pi = \int\limits_\Omega \pi\, d\Omega,
\end{equation*}
\begin{equation}\label{eq:en:4th}
  \begin{aligned}
    \pi
    & = \pi(\Phi,\phi,\nabla\phi,\Delta\phi)  \\
    & = -\frac{1}{2}\epsilon[\phi]\nabla\Phi\dpr\nabla\Phi
    + \Gamma \frac{1-f(\phi)}{l^2}
    + \frac{\Gamma}{4}\nabla\phi\dpr\nabla\phi
    + {\alpha}\frac{\Gamma l^2}{8}(\Delta\phi)^2
    + {\beta}\frac{1}{p}{\Gamma}{l^{p-2}} \|\nabla \phi\|^p,
  \end{aligned}
\end{equation}
where~$p\geq 4$ is even natural number, 
$\|\cdot\|$ is the standard Euclidean norm in~$\mathbb{R}^3$,
$ \alpha, \beta \geq 0 $ are positive real parameters,
unequal to zero simultaneously, i.e.,~$\alpha + \beta> 0 $.

The complete system of equations describing evolution of the electric
potential~$\Phi$ and order parameter~$\phi$, has the
form~\eqref{eq:el} or, in the particular case under consideration,
\begin{subequations}\label{eq:pp:4th}
  \begin{gather}
    \label{eq:pp:4th:a}
    \nabla \dpr (\epsilon(\phi)\nabla \Phi) = 0,\\
    \label{eq:pp:4th:b}
    \frac{1}{m}\dudx{\phi}{t}
    = \frac{1}{2}\epsilon'(\phi)\nabla\Phi\dpr\nabla\Phi
    + \frac{\Gamma}{l^2}f'(\phi)
    + \frac{\Gamma}{2}\Delta\phi
    - {\alpha}\frac{\Gamma l^2}{4}\Delta^2\phi
    + {\beta}\Gamma l^{p-2}\nabla\dpr\left( \|\nabla\phi\|^{p-2} \nabla\phi \right) . 
  \end{gather}
\end{subequations}

A priori, one cannot state which one of the regularizing
terms in the expression~\eqref{eq:en:4th} and~\eqref{eq:pp:4th}
is preferable either form thermodynamic or numerical point of views~--- or both are needed.
We only note that~$\alpha> 0 $
and~$\beta=0$ leads to the linear biharmonic term
in~\eqref{eq:pp:4th:b} which poses certain problems in numerical
approximations.  As will be shown below, this makes solution
of~\eqref{eq:pp:4th:b} to be smooth at  points of~$\Lambda$.
The case~$ \alpha = 0 $, $\beta> 0 $ leads to $p$-Laplacian term
in~\eqref{eq:pp:4th:b} which is nonlinear, but of the second order.

A consequence of the second equation of~\eqref{eq:pp:4th}
is the following model equation describing the evolution of the order parameter:
\begin{equation}\label{eq:pr:3d}
 \frac{1}{m} \dudx{\phi}{t}
 + {\alpha}\frac{\Gamma l^2}{4}\Delta^2\phi
 - {\beta}\Gamma l^{p-2}\nabla\dpr\left( \|\nabla\phi\|^{p-2}\nabla\phi\right)
  - \frac{\Gamma}{2}\Delta\phi
  - \frac{\Gamma}{l^2}f'(\phi)  = 0.
\end{equation}

For~$ \beta = 0 $ this equation is widely known as Swift-Hohenberg equation.
It is the basic equation in phase-field crystal models.
These models are under active development, starting from the fundamental
work~\cite{elder_2002}, see review in~\cite{provatas_2007, emmerich_2012}.
Note that the original model proposed
in~\cite{elder_2002} contains only a biharmonic term. Examples of Swift-Hohenberg type
models which have both the biharmonic and the $p$-Laplacian terms
are given in~\cite{golovin_2003}, see also references therein.

Note that derivation of the Swift-Hohenberg equations known in the
literature is physically justified, and is not mechanistic in the
sense of section~\ref{sec:intro}: the term, proportional
to~$ \Delta^2 \phi $ in the expression for the energy for these models appears
from physical considerations related to the form of the free energy
function of the corresponding atomic system, see, e.g.,~\cite{emmerich_2012, elder_2007}. 

Finally, let us note that:
\begin{itemize}
\item An example of a high-order diffuse boundary model for fractures
  is given in~\cite{borden_2014}.  The motivation for using such a
  model in the specified work is to improve the smoothness properties
  of of the solution and, as a consequence, the computational
  properties of the isogeometric finite element method used for
  numerical solution of the problem.  Thus, the generalizations made
  in~\cite{borden_2014} are intended to be rather technical and is
  not related to the correctness of the diffuse interface model.

\item As noted in~\cite{borden_2014}, for diffuse boundary models of
  higher order (with biharmonic term) for fractures in an elastic medium,
  rigorous and complete results concerning $\Gamma$-limit of  these models
  are not known. In other words, unlike for the classic diffuse interface
  models for fractures, it is not proven for high-order models, that
  they approximate the classic Grifftis model of fracture.
  Nevertheless, the results of numerical calculations suggest that
  this issue is rather technical.
\end{itemize}

\section{Numerical experiments}

The system of equations~\eqref{eq:pp:4th} is nonlinear and of the high order. 
So it is difficult to predict it's properties, especially analytically.
To get insight in it's properties we provide below number of numerical experiments.

Consider equation~\eqref{eq:pr:3d} in the stationary case:
\begin{equation}\label{eq:pr:3d:a}
  {\alpha}\frac{\Gamma l^2}{4}\Delta^2\phi
  - {\beta}\Gamma l^{p-2}\nabla\dpr\left( \|\nabla\phi\|^{p-2}\nabla\phi\right)
  - \frac{\Gamma}{2}\Delta\phi
  - \frac{\Gamma}{l^2}f'(\phi)  = 0.
\end{equation}
Let~$x_i = L \tilde{x}_i$, $i=\overline{1,3}$ with~$L$~being the characteristic length,
$\Tilde{x}_i$ is dimensionless coordinates.
The dimensionless form of this equation is:
\begin{equation}\label{eq:pr:3d:b}
  {\alpha}\frac{ (l/L)^4}{4}\Tilde{\Delta}^2\phi
  - {\beta} (l/L)^{p}\Tilde\nabla\dpr\left( \|\Tilde{\nabla}\phi\|^{p-2}\Tilde{\nabla}\phi\right)
  - \frac{(l/L)^2}{2}\Tilde{\Delta}\phi
  - f'(\phi)  = 0,
\end{equation}
where~$\Tilde{\nabla}$ and~$\Tilde\Delta$ are Hamilton and Laplace operators in dimensionless coordinates. 

If characteristic length is chosen as~$L=l$, one obtains:
\begin{equation}\label{eq:pr:3d:c}
  {\alpha}\frac{1}{4}{\Delta}^2\phi
  - {\beta} \nabla\dpr\left( \|{\nabla}\phi\|^{p-2}{\nabla}\phi\right)
  - \frac{1}{2}{\Delta}\phi
  - f'(\phi)  = 0.
\end{equation}
In what further only dimensionless equations will be considered with the
tilde symbol omitted.

In this section we will study dependence of numerical solutions of equation~\eqref{eq:pr:3d:c} parameters values.
The following properties if solution of the equation~\eqref{eq:pr:3d:c}
are of the interest
\begin{itemize}
\item  the solution range must be a segment~$\phi\subset[0,1]$;

\item distribution of phase-field must be monotonic;

\item numerical convergence must be observed with reasonable refinement of the computational grid;

\item phase-field function values closed to zero has to be concentrated in the sufficiently small
  neighbourhood of~$r=0$,~---  
    so that the solution goes to the asymptotic~$\phi\to 1$, $\partial\phi/\partial r\to 0$ when~$r\to+\infty$
pretty fast.

\end{itemize}
One must notice that these properties are not obvious for the solution of considered equation.

Numerical simulations were performed for the one dimensional axisymmetric problem.
We assume that:
\begin{itemize}
\item breakdown channel is aligned along axis~$\mathcal{O}z$ of cylindrical coordinate system~$\mathcal{O}r{\theta}z$;
  
\item solution of the equation~\eqref{eq:pr:3d:c} doesn't depend on
  $z$-coordinate and is axisymmetric, so~$\partial\phi/\partial z = 0$ and~$\partial\phi/\partial \theta = 0$.  
\end{itemize}

In this case, solution of equation~\eqref{eq:pr:3d:c} depends only on coordinate~$r$
defined in cylindrical domain
\[
  \Omega = \left\{ z\in\mathbb{R},\; \theta\in(0,2\pi],\; r \in (0,R \right) \},
\]

In the considered axisymmetric case the equation~\eqref{eq:pr:3d:c} reads:
\begin{equation}\label{eq:pr:1d}
  {\alpha} \frac{1}{4}\frac{1}{r}\dudx{}{r}\left(r \dudx{}{r}\left[\frac{1}{r}\dudx{}{r}\left(r \dudx{\phi}{r}\right)\right]\right)
  - {\beta}
  \frac{1}{r}\dudx{}{r}\left( r \left[ \dudx{\phi}{r}  \right]^{p-2}  \dudx{\phi}{r} \right)
  - \frac{1}{2}\frac{1}{r}\dudx{}{r}\left(r \dudx{\phi}{r}\right)
  - f'(\phi) = 0.
\end{equation}
The boundary conditions are
\begin{alignat*}{2}
  &\text{for } r = 0:& \quad & \phi(\vcx)  = 0,\; \left.\dudx{\phi}{r}\right|_{r=0} = 0, \\
  &\text{for } r = R:& \quad & \phi(\vcx)  = 1,\; \left.\dudx{\phi}{r}\right|_{r=R} = 0,
\end{alignat*}
when~$\alpha \neq 0$ and 
\begin{alignat*}{2}
  &\text{for } r = 0:& \quad & \phi(\vcx)  = 0,\\
  &\text{for } r = R:& \quad & \phi(\vcx)  = 1.
\end{alignat*}
when~$\alpha = 0$.

For numerical solution of the problem, a finite  difference method was
used.

Consider a non-uniform computational mesh with~$N$ nodes
defined in~$\Omega = [0,R]$:
\[
  0 = r_0 < r_1 < \ldots < r_{N-2} < r_{N-1} = R.
\]
Let~$\omega_{i+1/2} = [r_i,r_{i+1}]$ be mesh cells with 
centers denoted by~$r_{i+1/2}$, 
\[
  r_{i+1/2} = (r_i + r_{i+1})/2, \quad i = \overline{0,N-2}.
\]
The cell centers form a dual mesh with cells~$\omega_i = [r_{i-1/2},r_{i+1/2}]$,
$i=\overline{1,N-1}$.
In what further let~$\Delta r_{i+1/2} = |\omega_{i+1/2}| = r_{i+1} - r_i$,
$i=\overline{0,N-2}$ be a mesh step size for the primary mesh,
$\Delta r_{i} = |\omega_{i}| = r_{i+1/2} - r_{i-1/2}$,
$i = \overline{0,N-2}$ be be a mesh step size for the dual mesh.

The values of~$\phi$ related to the nodes of primary and dual meshes
are denoted as~$\phi_i$ and~$\phi_{i+1/2}$, respectively.
The correposnding mesh function is denoted as~$\phi_h = (\phi_0,\phi_1,\ldots,\phi_{N-1})$.

At the nodes of the primary mesh,
the finite-difference approximation of the equation~\eqref{eq:pr:1d}
is defined at the nodes of the primary mesh and read:
\begin{gather}
  \label{eq:fd}
  \alpha \frac{1}{4}\Delta^2_h\phi_h
  - \beta\Delta_{p,h}\phi_h
  - \Delta_h\phi_h
  - f'_{i}
  = 0,
\end{gather}
with the discrete Laplace operator~$\Delta^2_h\phi_h$ defined as 
\[
  \left[\Delta_h\phi_h\right]_i =
      \frac{1}{r_i}\cdot \frac{1}{\Delta r_i}
      \left[
        r_{i+1/2} \frac{\phi_{i+1} - \phi_i}{\Delta r_{i+1/2}} -
        r_{i-1/2} \frac{\phi_{i} - \phi_{i-1}}{\Delta  r_{i-1/2}} 
      \right],\quad i = \overline{1,N-2}.
\]
discrete~$p$-Laplacian  defined as:
\begin{multline*}
  \left[\Delta_{p,h}\phi_h\right]_i = 
  \frac{1}{r_i}\frac{1}{\Delta r_i}\biggl(
  r_{i+1/2} \left| \frac{\phi_{i+1}-\phi_{i}}{\Delta r_{i+1/2}}  \right|^{p-2}  \frac{\phi_{i+1}-\phi_{i}}{\Delta r_{i+1/2}} -  \\
  r_{i-1/2} \left| \frac{\phi_{i}-\phi_{i-1}}{\Delta r_{i-1/2}}  \right|^{p-2}  \frac{\phi_{i}-\phi_{i-1}}{\Delta r_{i-1/2}}
  \biggr),
  \quad i = \overline{1,N-2},
\end{multline*}
and the discrete biharmonic operator~$\Delta^2_h\phi_h$ defined as:
\[
  \left[\Delta^2_h\phi_h\right]_i  = \left[  \Delta_h \circ \Delta_h\phi_h\right]_i,\quad i = \overline{2,N-3}.
\]

In the expression above the term~$f'_i$ is defined as 
\[
  f'_i = f'(\phi_i).
\]

Let's denote finite-difference approximation of spatial derivatives by the index~$h$,
e.g.,~$\partial_h \phi_h/\partial_h r$ is approximation to the~$\partial\phi/\partial r$. 

For~$\alpha \neq 0$ the difference equation~\eqref{eq:fd} is defined at the nodes~$i=\overline{2,N-3}$,
i.e., in all primary mesh nodes, except 
the first two and the last two nodes, where boundary conditions are defined:
\begin{alignat*}{2}
  &\text{for } r = 0:& \quad & \phi_0  = 0, \left[\dhudhx{\phi_h}{r}\right]_0 = 0,
  \\
  &\text{for } r = R:& \quad & \phi_{N-1}  = 1,\left[\dhudhx{\phi_h}{r}\right]_{N-1} = 0.
\end{alignat*}
Here the difference derivative~$\left[{\partial_h\phi_h} / \partial_h{r}\right]_0$ is approximated  with the second order
using three-point stencil, 
\[
\left[\dhudhx{\phi_h}{r}\right]_0 =  
-\frac{2\Delta r_{1/2}+\Delta r_{1+1/2}}{\Delta r_{1/2}(\Delta r_{1+1/2}+\Delta r_{1/2})}\phi_0+\frac{\Delta r_{1/2}+\Delta r_{1+1/2}}{\Delta r_{1/2}\Delta r_{1+1/2}}\phi_1-\frac{\Delta r_{1/2}}{(\Delta r_{1/2}+\Delta r_{1+1/2})\Delta r_{1+1/2}}\phi_2.
\]
For a uniform mesh with mesh step~$\Delta r$ this expression simplifies to:
\[
  \left[\dhudhx{\phi_h}{r}\right]_0 = \frac{-3\phi_0+4\phi_1-\phi_2}{2\Delta r}.
\]
The derivative~$\left[\partial_h{\phi} / \partial_h{r}\right]_{N-1}$ is approximated in the same way.

For~$\alpha = 0$, the difference equation~\eqref{eq:bc} is defined at the nodes~$i=\overline{1,N-2}$,
except only the first and the last node of the mesh where the boundary conditions are defined,
\begin{alignat*}{2}
  &\text{for } r = 0:& \quad & \phi_0  = 0,\\
  &\text{for } r = R:& \quad & \phi_{N-1}  = 1.
\end{alignat*}

The constructed difference scheme is nonlinear, i.e., it has the form of a system of nonlinear
algebraic equations for solution's nodal values.
There are different ways to solve it. We employed the Newton's method with
iterations performed until the value of 2-norm of residual decreases in at least~$\varepsilon = 10^{-6}$
times. 

\subsection{Series of calculations 1.}
In this series of calculations, we demonstrate qualitative dependency
of the solution of~\eqref{eq:pr:1d} on the values of parameters~$\alpha$ and~$\beta$,
\[
  (\alpha,\beta)\in \{0, 10^{-2}, 10^{-1}, 1\}^2.
\]

\begin{figure}[tp!]
  \centering
  \subfloat[$\alpha=0$, $\beta=0$     ]{ \includegraphics[width=0.225\textwidth]{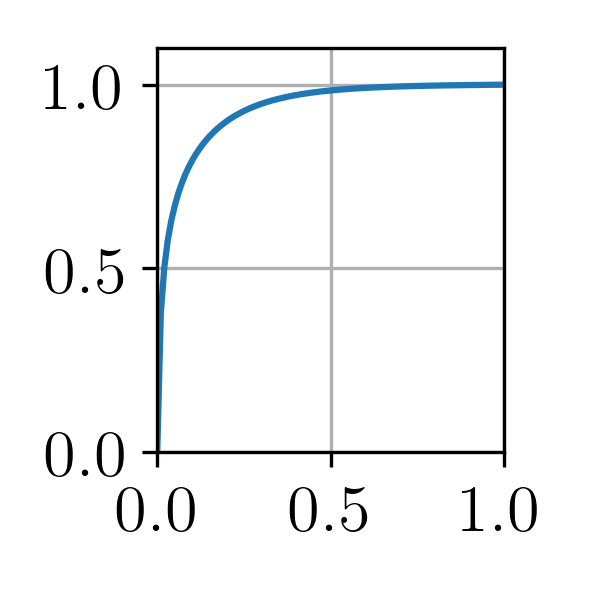} }
  \hfill
  \subfloat[$\alpha=0$, $\beta=10^{-2}$]{ \includegraphics[width=0.225\textwidth]{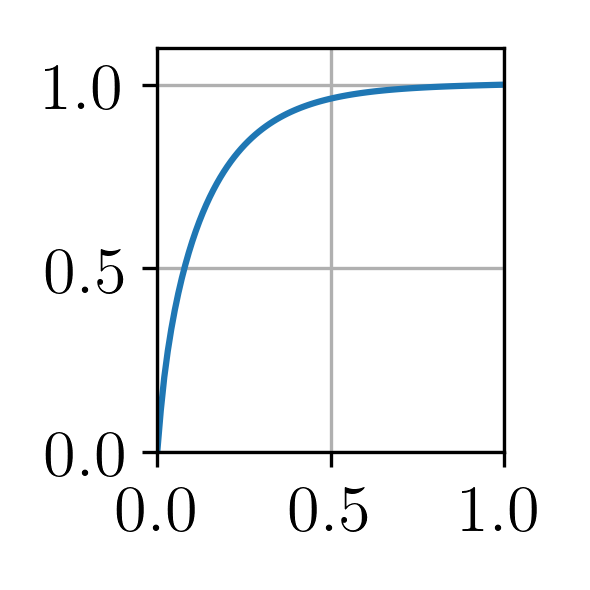} }
  \hfill
  \subfloat[$\alpha=0$, $\beta=10^{-1}$]{ \includegraphics[width=0.225\textwidth]{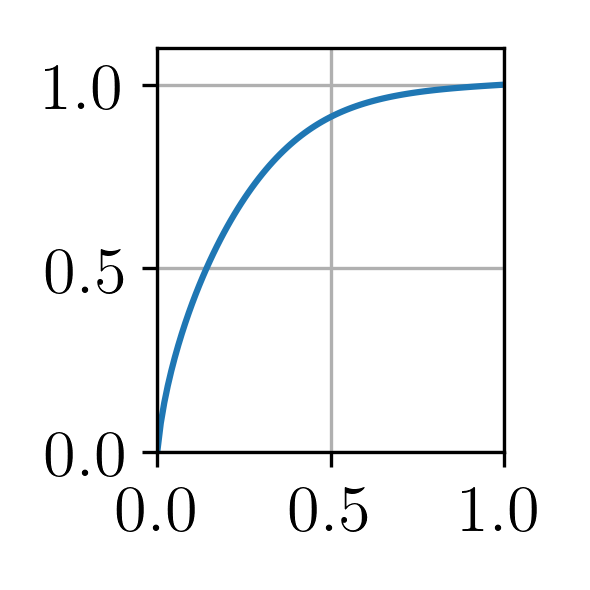} }
  \hfill
  \subfloat[$\alpha=0$, $\beta=10^{0}$ ]{ \includegraphics[width=0.225\textwidth]{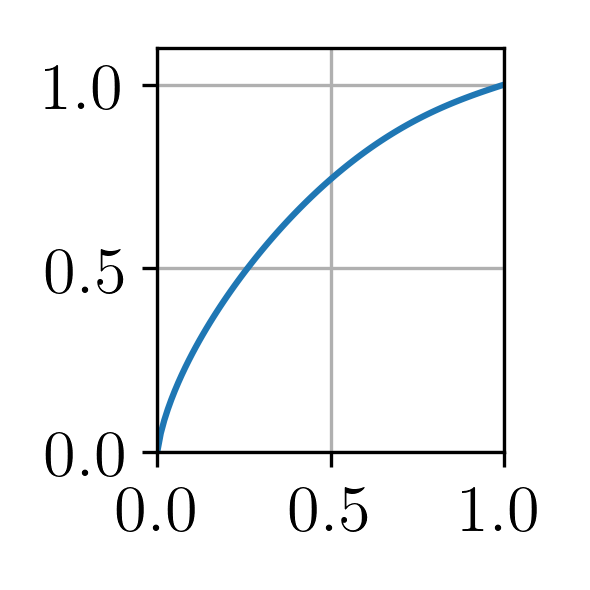} }
  \\
  %
  \subfloat[$\alpha=10^{-2}$, $\beta=0$      ]{ \includegraphics[width=0.225\textwidth]{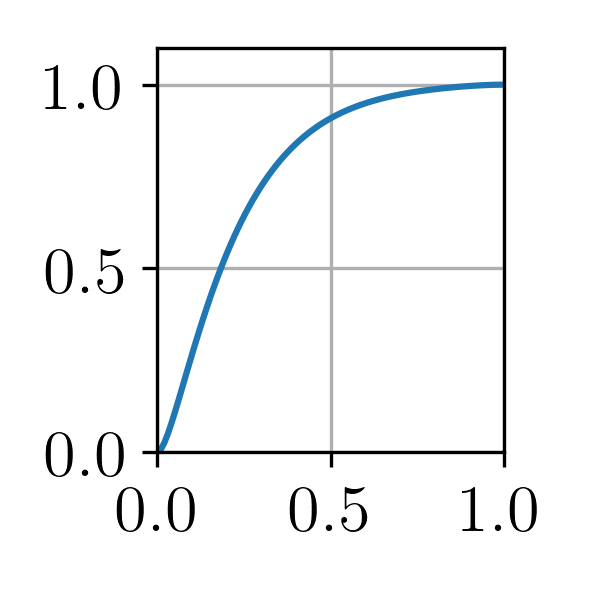} }
  \hfill
  \subfloat[$\alpha=10^{-2}$, $\beta=10^{-2}$]{ \includegraphics[width=0.225\textwidth]{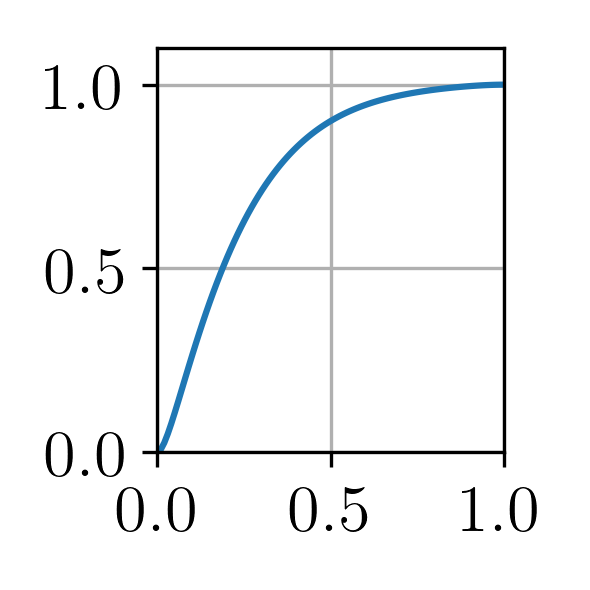} }
  \hfill
  \subfloat[$\alpha=10^{-2}$, $\beta=10^{-1}$]{ \includegraphics[width=0.225\textwidth]{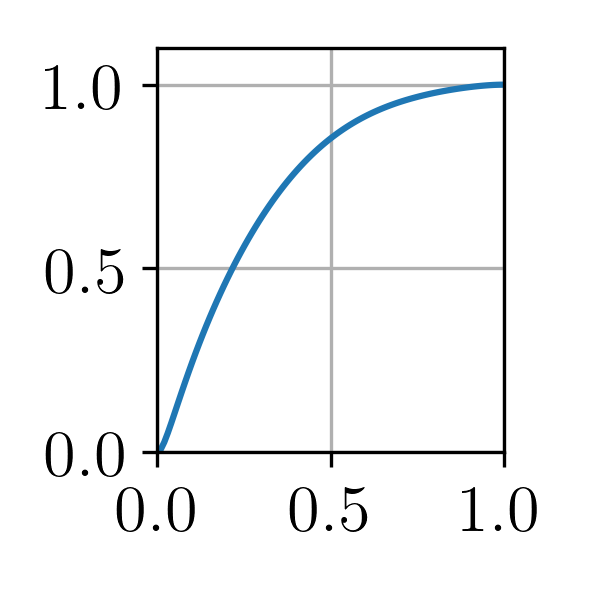} }  
  \hfill
  \subfloat[$\alpha=10^{-2}$, $\beta=10^{0}$]{ \includegraphics[width=0.225\textwidth]{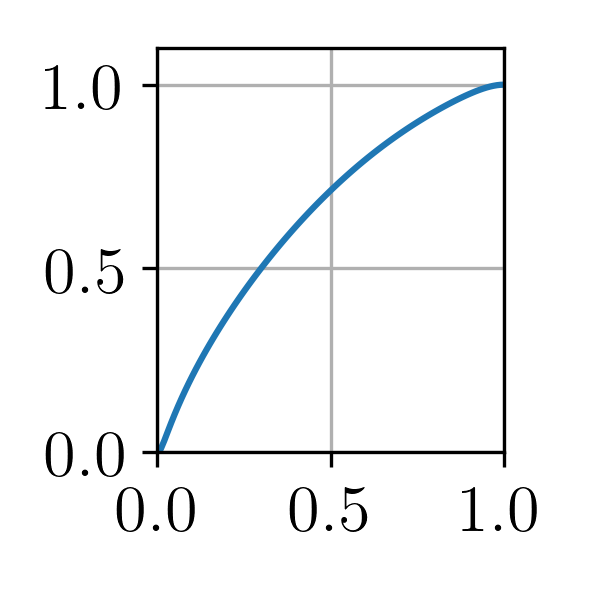} }    
  \\
  \subfloat[$\alpha=10^{-1}$, $\beta=0$      ]{ \includegraphics[width=0.225\textwidth]{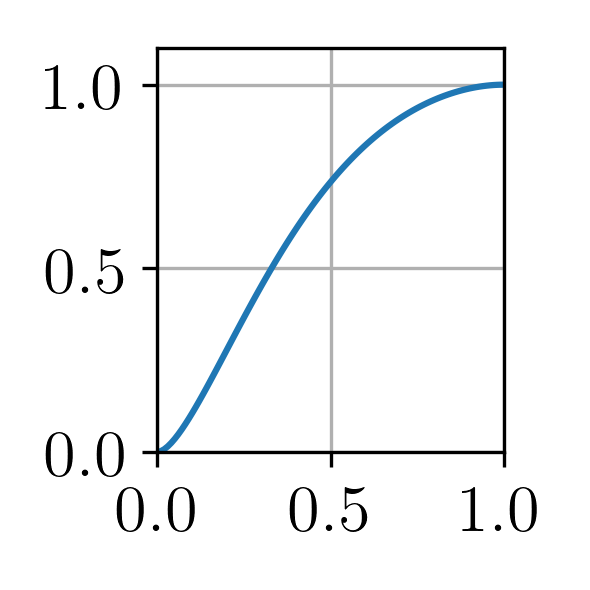} }
  \hfill
  \subfloat[$\alpha=10^{-1}$, $\beta=10^{-2}$]{ \includegraphics[width=0.225\textwidth]{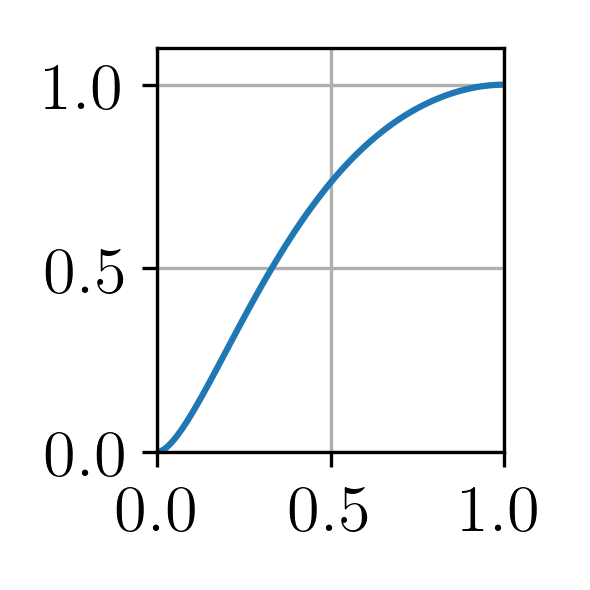} }
  \hfill
  \subfloat[$\alpha=10^{-1}$, $\beta=10^{-1}$]{ \includegraphics[width=0.225\textwidth]{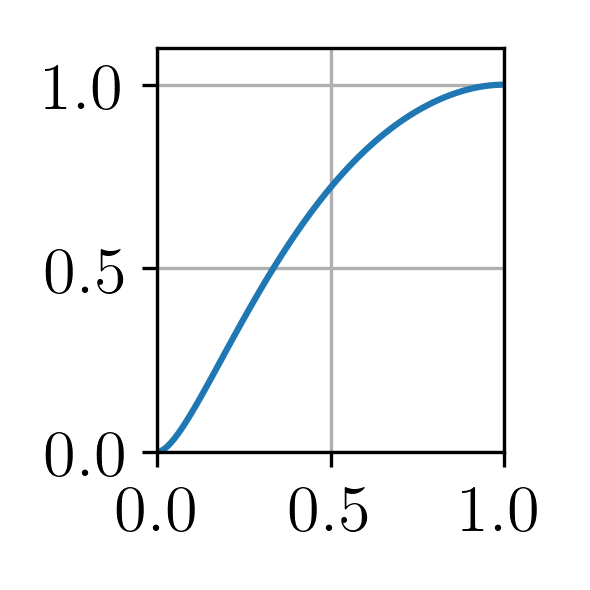} }    
  \hfill
  \subfloat[$\alpha=10^{-1}$, $\beta=10^{0}$]{ \includegraphics[width=0.225\textwidth]{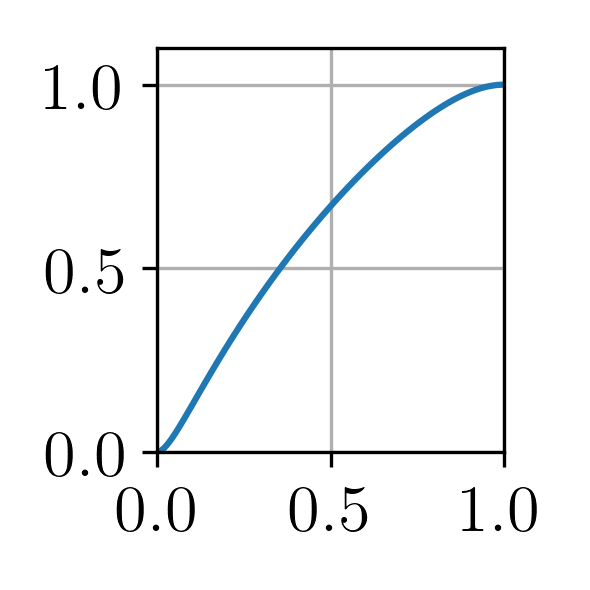} }
  \\
  \subfloat[$\alpha=10^{0}$, $\beta=0$      ]{ \includegraphics[width=0.225\textwidth]{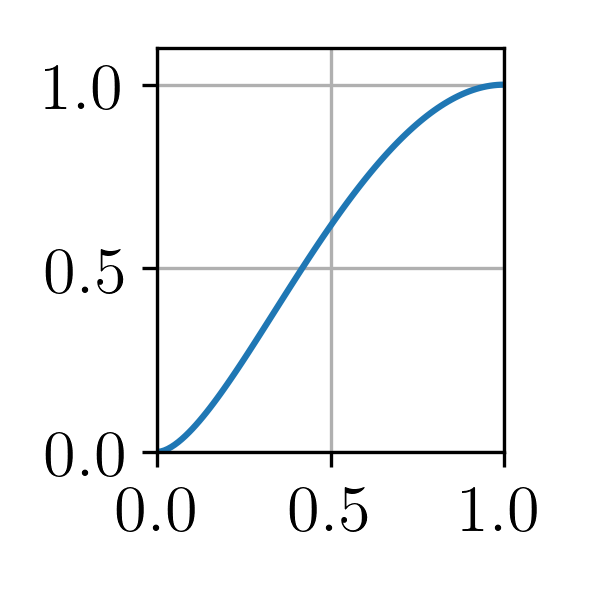} }
  \hfill
  \subfloat[$\alpha=10^{0}$, $\beta=10^{-2}$]{ \includegraphics[width=0.225\textwidth]{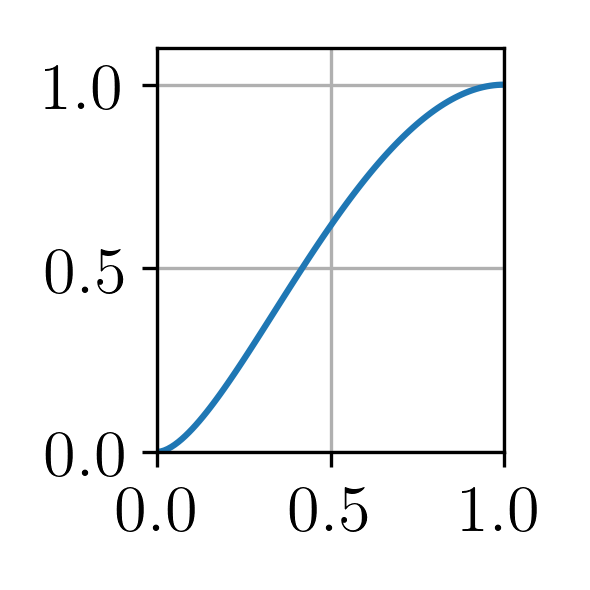} }
  \hfill
  \subfloat[$\alpha=10^{0}$, $\beta=10^{-1}$]{ \includegraphics[width=0.225\textwidth]{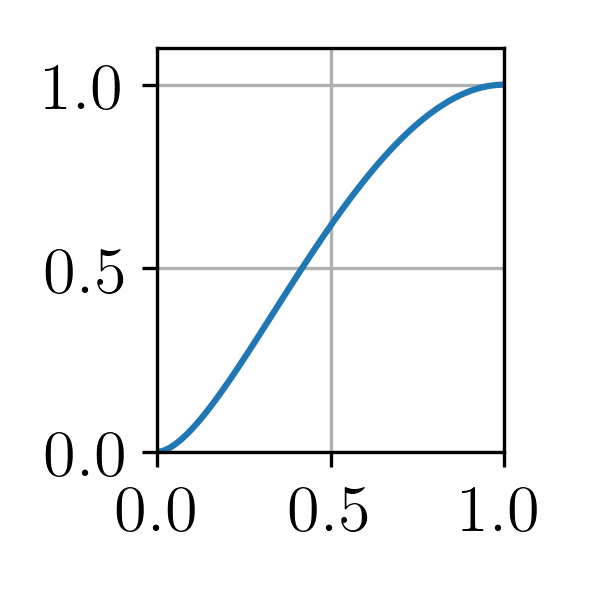} }
  \hfill
  \subfloat[$\alpha=10^{-0}$, $\beta=10^{0}$]{ \includegraphics[width=0.225\textwidth]{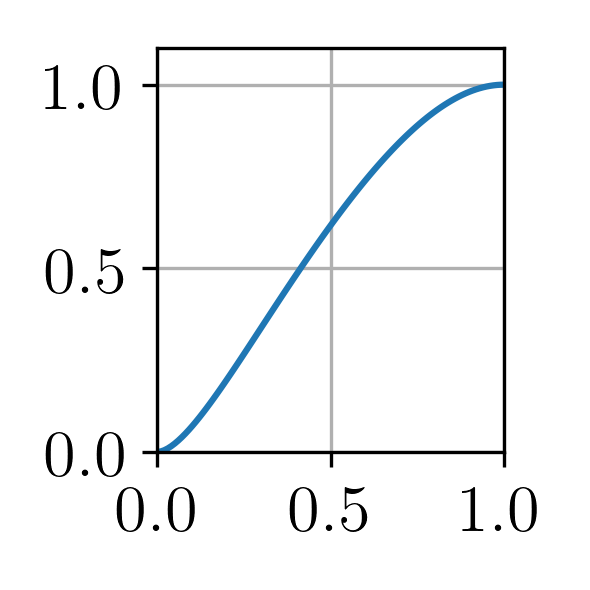} }

  \caption{
    Phase field distribution for~$(\alpha,\beta)\in \{0, 10^{-2}, 10^{-1}, 1\}^2$.}\label{fig:ab}
  
\end{figure}

Computational domain is of the radius~$R=1$ in dimensionless coordinates.
The uniform mesh with~$N=100$ nodes was used. In figure~\ref{fig:ab},
the solutions of the problem are shown.
We observed that for all values of parameters the corresponding
solution is monotonic.
Let us note that this result is not general~---
e.g., it is known that for generalized Fisher-Kolmogorov equation
(which is equation~\eqref{eq:pr:1d} with~$\beta=0$, the mentioned
properties of solution strongly depends on the values of parameter
$\alpha$ and the choice of the function~$f$, see, e.g.,~\cite{bonheure_2016}). 

The rows of the table in the figure~\ref{fig:ab}
correspond to the constant values of~$\alpha$,
its columns~--- to the constant values of~$\beta$.
The top left plot correspond to the incorrect, formal, solution
of the difference scheme.
\begin{figure}[ht!]
  \centering
  
  \includegraphics[width=0.45\textwidth]{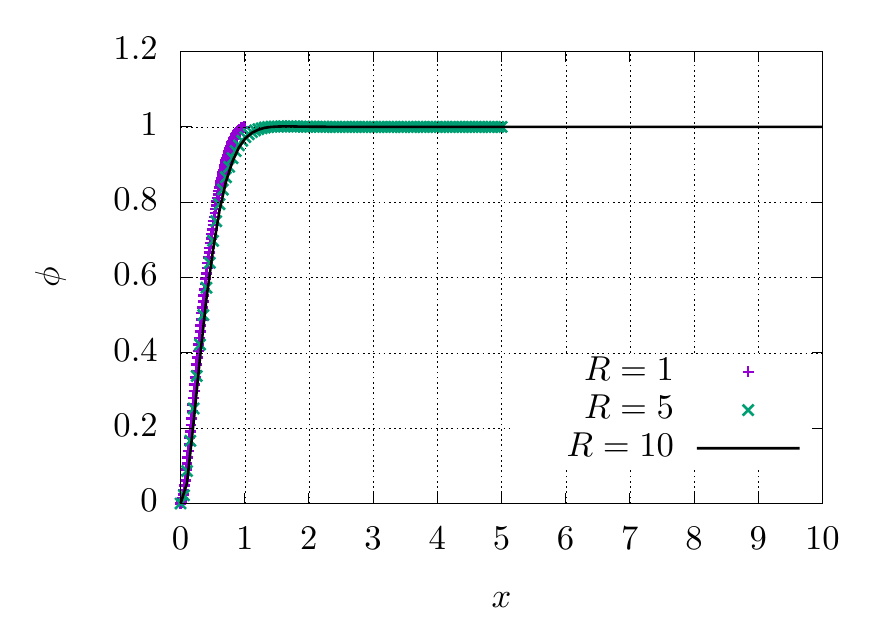}
  \includegraphics[width=0.45\textwidth]{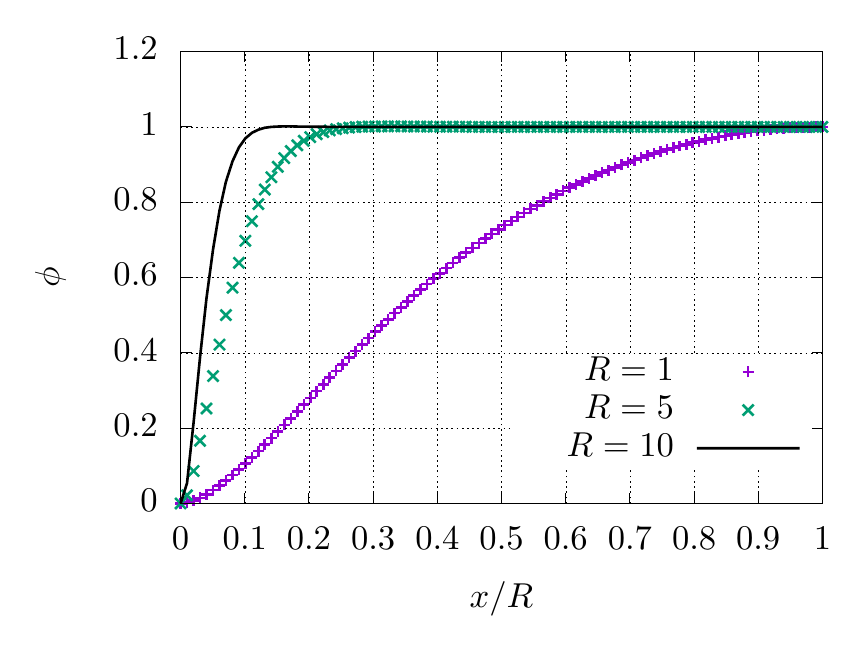}
  \caption{Phase field distribution for~$R=1,5,10$.
    Solid line is solution for~$R=10$, points~--- for~$R=1,5$.}\label{fig:r}
  
\end{figure}

\subsection{Series of calculations~2.}
In this series of calculations we show that the solution of the problem~~\eqref{eq:pr:1d}
has effectively finite support as the radius of the domain increases.

Uniform mesh with~$N=100$ nodes was used. The
parameters of the equation~\eqref{eq:pr:1d} were chosen as~$\alpha=0.1$, $\beta=0.01$.
In the figure~\ref{fig:r} solutions of the problem for~$R=1,5,10$ are shown.

\subsection{Series of calculations 3.}
In this series of calculations, the mesh convergence
was studied for the original equation, presented in~\cite{pitike_2014}
(which corresponds to~$\alpha=\beta=0$ here)~--- and for the corrected one,
introduced in the current work (with~$\alpha=\beta=0.1$).

Three meshes were chosen: two uniform meshes with the number of nodes~$N=100$ and~$N=1000$,
and an two adaptive meshes with logarithmic distribution of nodes.
In the last case the nodal coordinates are defined
as~$r_i = \left(10^{-8}\right)^{(N-1-i)/N}$, $i=\overline{0,N-1}$ with number of nodes~$N=100$.
In this case the minimal mesh step value at~$r=0$ is~$\Delta\approx 10^{-8}$.
The calculation domain is of the radius~$R=1$ in dimensionless coordinates.

\begin{figure}[t!]
  \centering
  \includegraphics[width=0.45\textwidth]{figs/alpha_0_beta_0_varmesh.pdf}
  \includegraphics[width=0.45\textwidth]{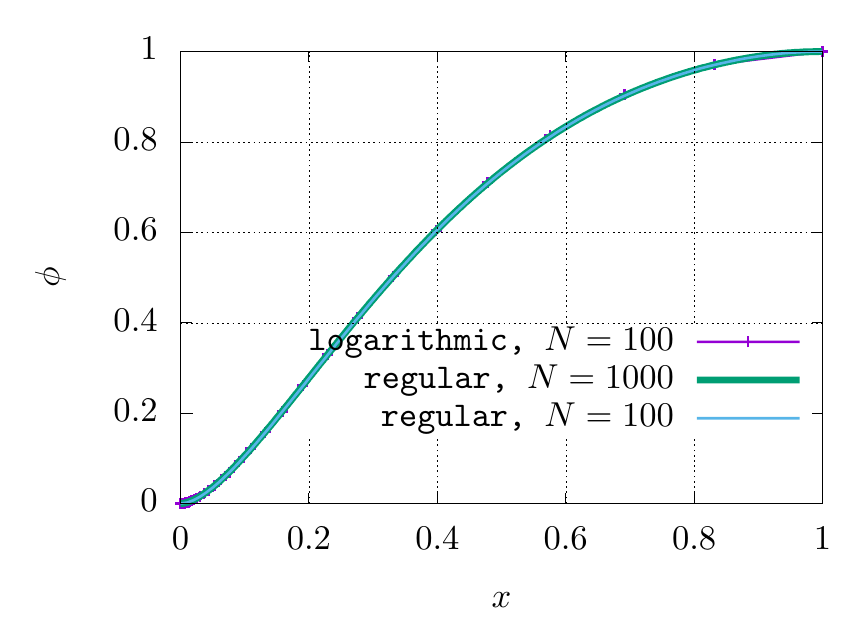}
  \caption{Phase field distribution for different grids for set of parameters on the left:~$\alpha=0.0$, $\beta=0.0$; on the right:~$\alpha=0.1$, $\beta=0.01$}\label{fig:varmesh}
\end{figure}

In the figure~\ref{fig:varmesh} it is clearly seen that
for~$\alpha = \beta = 0.0$ when the mesh is refined, the numerical
solution approaches function~$\phi = 0$
in all points of~$\Omega$~---
except the point~$r=0$ where~$\phi = 1$ and
$\partial\phi/\partial r\to\infty$.
In this case, the form of the numerical solutions
changes significantly upon mesh refinement.  That is, the numerical
solution of the problem posed in~\cite{pitike_2014} tends to the
distribution of the order parameter with an infinitely small
interface between the media, which is in consistency with the
theoretical analysis of section~\ref{sec:analysis}.  In particular,
this means that the ``thickness'' of the diffuse boundary does not
match the parameter~$l$ and is never resolved by mesh.

For the regularized model mesh  convergence is clearly observed even with relatively modest number of mesh points. 
The calculation results demonstrate mesh convergence and reflect the fact that the ``thickness'' of the diffuse boundary 
is a parameter of the model~--- and not the numerical  artifact of the computational algorithm
used to solve an ill-posed boundary value problem.

\section{Conclusions}

In this paper we shows that the diffuse interface electric breakdown
model suggested in~\cite{pitike_2014} is not completely correct from
mathematical viewpoint and based on a wrong assumptions on the expression for the
free energy. The source of the error appears to be a formal generalization
of the diffuse fracture models to the case of the (diffuse)
breakdown channel. This generalization does not take into account that
codimension of the fracture mid-surface and breakdown channel are
different.

Note that in~\cite{pitike_2014}, the simulation results are presented,
which can be characterized as quite reasonable. This does not
contradict the theoretical considerations, disccussed above. In
practice, the inconsistency of the model can be noticeable only when
studying the numerical convergence of the discrete approximations of
the model's equations. Such results are not presented in any of the
papers cited above, see~\cite{pitike_2014, hong_2015, cai_2017a,
cai_2017b, cai_2019a, cai_2019b, cai_2019c}. Most of these papers
explicitly state that the simulations were carried out using
commercial software.

From the theoretical viewpoint, one can expect that using an incorrect
expression for energy will result in inexistance of the $\Gamma$-limit
of the diffuse interface model and the corresponding sharp interface
counterpart.

Finally, the main, up to the authors opinion, conclusion that can be
drawn from this work is that, most likely, the use of high-order
(``high'' in the sense of ``number of derivatives'' or in the sense
of ``power of derivatives'')
diffuse boundary models is the~\emph{necessary} when a diffuse object is,
effectively, an object of codimension~2 or~3.

Although there are known (cited above) papers in which high-order
diffuse interface models are considered,~--- their use did not have
the character of a fundamental necessity. In other words, they are
only quantitatively improved previously known 
models. For example, in the cited above work~\cite{borden_2014}, the
higher order model is considered as a mean to improve accuracy of
isogeometric solvers.

In the case of the diffuse interface electric breakdown model
considered here, the use of high-order model is a prerequisite for
their mathematical correctness.


\end{document}